\PassOptionsToPackage{numbers,sort&compress}{natbib} 

\documentclass[11pt]{article}

\usepackage[margin=1in]{geometry}
\usepackage[T1]{fontenc}
\usepackage[utf8]{inputenc}
\usepackage{lmodern}
\usepackage{microtype}
\usepackage{url}         
\usepackage{amsmath,amssymb,amsfonts}

\usepackage{graphicx}
\usepackage{booktabs}
\usepackage{longtable}
\usepackage{array}
\usepackage{makecell}
\usepackage{enumitem}
\newcolumntype{L}[1]{>{\raggedright\arraybackslash}p{#1}}
\newcolumntype{C}[1]{>{\centering\arraybackslash}p{#1}}

\usepackage{tikz}
\usetikzlibrary{arrows.meta,positioning,calc,matrix}

\usepackage[numbers,sort&compress]{natbib}          
\usepackage[hidelinks]{hyperref}                    
\usepackage[nameinlink,noabbrev,capitalise]{cleveref} 

\usepackage{authblk}

\DeclareUnicodeCharacter{200B}{}
\DeclareUnicodeCharacter{2212}{$-$}
\DeclareUnicodeCharacter{0394}{$\Delta$}
\DeclareUnicodeCharacter{2248}{$\approx$}
\DeclareUnicodeCharacter{2265}{$\ge$}

\title{Scaling Multilingual Semantic Search in Uber Eats Delivery}

\author[1]{Bo Ling\thanks{\texttt{bo.ling@uber.com}; personal: \texttt{lbwavebo@gmail.com}}}
\author[1]{Zheng Liu\thanks{\texttt{zheng.liu@uber.com}; personal: \texttt{z0liu@outlook.com}}}
\author[1]{Haoyang Chen\thanks{\texttt{haoyang.chen@uber.com}; personal: \texttt{haoychen@outlook.com}}} 
\author[1]{Divya Nagar\thanks{\texttt{divya.nagar@uber.com}; personal: \texttt{divn2661@gmail.com}}}
\author[1]{Luting Yang\thanks{\texttt{luting.yang@uber.com}}}
\author[1]{Mehul  Parsana\thanks{\texttt{mehul.parsana@gmail.com}}}
\affil[1]{Uber Technologies Inc.}

\date{} 

\begin{document}
\maketitle

\vspace{-0.75em}
\noindent\textbf{Comments:} 16 pages, 11 tables, 1 figure. Planned for submission to SIGIR, SIGIR eCom, or KDD 2026.
\vspace{0.75em}

\begin{abstract}
We present a production-oriented semantic retrieval system for Uber Eats that unifies retrieval across stores, dishes, and grocery/retail items. Our approach fine-tunes a Qwen2 two-tower base model using hundreds of millions of query-document interactions that were aggregated and anonymized pretraining. We train the model with a combination of InfoNCE on in-batch negatives and triplet-NCE loss on hard negatives, and we leverage Matryoshka Representation Learning (MRL) to serve multiple embedding sizes from a single model. Our system achieves substantial recall gains over a strong baseline across six markets and three verticals. This paper presents the end to end work including data curation, model architecture, large-scale training, and evaluation. We also share key insights and practical lessons for building a unified, multilingual, and multi-vertical retrieval system for consumer search.
\end{abstract}


\section{Introduction}
Search is a primary entry point for Uber Eats users, and retrieving relevant entities---whether they are stores, dishes, or grocery items---demands a sophisticated understanding of a user's intent beyond simple keyword matching. Historically, our systems relied on fragmented solutions, using either separate BERT-based models for each vertical \cite{karpukhin2020dpr} or a simple keyword-based lexical search \cite{robertson2009bm25}. This approach increased operational complexity, limited generalization, and left significant performance headroom untapped.

We address these limitations by introducing a single, unified, multilingual, and multi-vertical system. This system is built on a two-tower architecture, where a dedicated query encoder and a distinct document encoder---both based on the latest open source decoder-only LLM structures (Qwen) \cite{qwen2024report,alibaba2024gteqwen} ---map their respective inputs into a shared embedding space. This decoupling of the encoders, which are trained with unshared weights, allows for a single retrieval index and enables the offline pre-computation of document embeddings for efficient real-time serving \cite{huang2013dssm,shen2014cdssm,covington2016youtube,lu2020twinbert,huang2020facebook}. Our training strategy is two-fold: a domain adaptation phase using hundreds of millions of large-scale click and ``add to cart'' (ATC) logs, followed by a fine-tuning phase with high-quality relevance data mined by a large language model \cite{google2025gemini,nvidia2024nvembed,li2023gte}. To support various downstream applications with different latency and size requirements, we employ Matryoshka Representation Learning (MRL), which allows a single embedding to be truncated to various dimensions, enabling efficient cost--performance trade-offs \cite{kusupati2022mrl}. Additionally, we are exploring a model-based neural distance function to move beyond the limitations of parameterless cosine similarity \cite{zhai2023neuralretrieval,linkedin2024linr}.

\paragraph{Our key contributions are:}
\begin{itemize}[leftmargin=*]
\item \textbf{A Unified Two-Tower System:} We developed a single, scalable, and multilingual two-tower model with distinct query and document encoders that unifies retrieval across three distinct verticals: restaurants, grocery and retail.
\item \textbf{A Scalable Training Recipe:} We designed a robust training pipeline that combines InfoNCE with in-batch negatives for large-scale domain adaptation, with an optional triplet loss for fine-tuning on hard-mined examples.
\item \textbf{Flexible Embeddings with MRL:} We implemented Matryoshka Representation Learning, enabling a single model to support multiple embedding sizes for on-the-fly dimensional trade-offs.
\item \textbf{Comprehensive Evaluation:} Extensive offline evaluations across six markets and three verticals demonstrate significant recall gains over strong baselines.
\item \textbf{Productionization and Innovation:} Practical insights into ANN indexing, online serving, and an exploration of model-based distance functions.
\end{itemize}

\section{Related Work}
The field of semantic retrieval has seen a rapid evolution, moving from lexical matching methods to dense retrieval models that leverage contextual embeddings \cite{robertson2009bm25,huang2013dssm,shen2014cdssm,karpukhin2020dpr,lu2020twinbert,huang2020facebook}. This transition has been driven by the rise of powerful pre-trained language models (PLMs) and their ability to encode rich, semantic information \cite{devlin2019bert,liu2019roberta}. Our work builds upon and contributes to several key areas within this landscape.

\subsection{Two-Tower Architectures}

The two-tower (or dual-encoder) model has become a foundational architecture for large-scale retrieval systems, particularly in e-commerce and search \cite{huang2013dssm,shen2014cdssm,covington2016youtube,lu2020twinbert,huang2020facebook}. This architecture, consisting of separate encoders for the query and document, allows for the pre-computation of document embeddings \cite{karpukhin2020dpr,khattab2020colbert}. This decoupling is crucial for efficiency, as it enables real-time retrieval from a massive corpus using Approximate Nearest Neighbor (ANN) search, meeting the low-latency requirements of online serving and standard ANN backends (FAISS/HNSW/quantization; BM25/Lucene hybrid) \cite{johnson2019faiss,malkov2018hnsw,guo2020anisotropic,robertson2009bm25,lucene2025doc}. Unlike computationally expensive cross-encoder models that concatenate query and document inputs, two-tower models sacrifice the early query–document interactions and restrict the interaction to a fixed late stage (e.g., cosine or a small scoring head), making them the standard for the first stage of retrieval. Our system adopts this proven architecture, focusing on optimizing the individual towers and their training process for our specific domain. 

\subsection{LLMs as Embedding Models}

Recent research has shown that large language models (LLMs), originally trained for generative tasks, can be fine-tuned to create highly effective embedding models \cite{muennighoff2022mteb,li2023gte,google2025gemini,nvidia2024nvembed}. These models, by virtue of their massive pretraining on vast and diverse corpora, possess a rich "world knowledge" that surpasses older, smaller PLMs like BERT or MiniLM \cite{devlin2019bert,wang2020minilm}. The Gemini Embedding paper demonstrates this by leveraging the powerful Gemini LLM to create a state-of-the-art embedding model that excels across over one hundred tasks on the Massive Multilingual Text Embedding Benchmark (MMTEB) \cite{google2025gemini}. Similarly, the NV-Embed paper shows how a decoder-only LLM can be enhanced with new architectural designs (e.g., a latent attention layer for pooling) and a two-stage contrastive instruction-tuning procedure to achieve state-of-the-art performance on various retrieval and non-retrieval tasks  \cite{nvidia2024nvembed}; and GTE provides a multi-stage contrastive recipe for general-purpose text embeddings \cite{li2023gte}. All papers highlight the importance of using a modern, capable LLM backbone and a sophisticated training methodology. Our work aligns with this trend, selecting the gte-Qwen2 model as our foundation to address the specific challenges — multilingual and multi-vertical — in the Uber Eats domain.  

\subsection{Training Techniques for Dense Retrieval}
The quality of embeddings is highly dependent on the training data and loss function. InfoNCE loss with in-batch negative sampling has emerged as a highly effective and scalable training objective for dense retrieval \cite{vandenOord2018cpc,chen2020simclr,gao2021simcse}. This approach treats all other examples in a mini-batch as negative samples for each positive pair, providing a high volume of negatives without the need for complex, explicit negative mining. This technique is particularly well-suited for the initial domain adaptation phase with massive, noisy click data.
For fine-tuning and refining the model's performance, hard negative mining is a critical component \cite{schroff2015facenet}. This involves identifying challenging examples that the model struggles with, such as documents that are semantically similar but irrelevant. The NV-Embed paper leverages this by using curated hard negative examples to improve model performance. We extend this idea by using an LLM to create a high-quality, explicitly labeled dataset of hard positive and negative examples to fine-tune our model's boundaries. 

\subsection{Flexible Embedding Architectures}
The need for a single embedding model to serve multiple downstream applications with varying computational constraints is a growing challenge. Matryoshka Representation Learning (MRL) offers an elegant solution by training a single high-dimensional embedding where the initial prefixes are as informative as independently trained lower-dimensional models \cite{kusupati2022mrl}. This allows practitioners to trade off between accuracy and speed at inference time without the operational complexity of managing multiple models. Our implementation of MRL enables us to use a single model to support fast retrieval with lower-dimensional embeddings while also providing a high-quality, full-dimensional embedding for downstream ranking models. 

\subsection{Multi-Modal and Multi-Vertical Embeddings}

Training a single model to handle diverse types of documents (e.g., stores, dishes, grocery items) is a key challenge in large-scale e-commerce search \cite{huang2020facebook,alayrac2022flamingo}. Prior work often relied on separate models for each vertical, leading to fragmentation and maintenance overhead. Our approach addresses this by using a structured JSON input format, which allows the gte-Qwen2 model to implicitly understand the document type from the field names and content

\subsection{Learned Non-linear Scoring}

Beyond fixed cosine/dot-product scoring, recent work explores learned similarity heads and model-based retrieval on accelerators. Revisiting Neural Retrieval on Accelerators proposes mixture-of-logits similarity with a hierarchical indexer to scale non-MIPS scoring, while LinkedIn’s LiNR executes GPU full-scan/model-based retrieval at production scale—offering complementary design points to dual-encoder + ANN pipelines \cite{mitra2017duet,zhai2023neuralretrieval,linkedin2024linr}. 

\section{System Overview}
We have built a production-grade semantic search system for Uber Eats that provides a unified solution for retrieving stores, dishes, and grocery items. The system operates on a two-tower architecture where query embeddings are generated online~\cite{huang2013dssm,shen2014cdssm,huang2020facebook}, while document embeddings are pre-computed offline. This design enables fast, scalable retrieval by leveraging Approximate Nearest Neighbor (ANN) search on a multi-billion document corpus~\cite{johnson2019faiss,malkov2018hnsw,guo2020anisotropic} . The same embeddings are also used as features for downstream ranking models~\cite{covington2016youtube,huang2020facebook}, ensuring consistency across the search funnel.

\subsection{Training and Model Generation}

Our model is a two-tower deep neural network built on a \textbf{gte-Qwen2} backbone. We use \textbf{PyTorch}~\cite{paszke2019pytorch} with \textbf{Hugging Face Transformers}~\cite{wolf2020transformers}  for model development and \textbf{Ray}~\cite{moritz2018ray} for distributed training orchestration. For large-scale training, we utilize \textbf{PyTorch's Distributed Data Parallel (DDP)}  in conjunction with \textbf{DeepSpeed ZeRO-3}~\cite{rajbhandari2020zero}, which shards the optimizer, gradients, and model parameters to handle the immense size of the Qwen2 model. This setup, combined with mixed-precision training and gradient accumulation~\cite{micikevicius2018mixedprecision}, allows us to train on hundreds of millions of aggregated and anonymized data points efficiently. Each successful training run generates versioned artifacts---a unique ID for the query and document models (\textit{query\_model\_id}, \textit{doc\_model\_id}) and a shared team-specific ID (\textit{tte\_id})---which are meticulously tracked for reproducibility and deployment~\cite{rajbhandari2020zero,huang2020facebook}.

\subsection{Online Query Serving}
The online serving path is optimized for low latency. When a user enters a search query, a request is first normalized into a structured JSON blob containing the search term, country, and language. This JSON is then fed into the query tower to generate a semantic embedding. To balance retrieval accuracy and serving latency, we leverage the \textbf{Matryoshka Representation Learning (MRL)} property~\cite{kusupati2022mrl}  of our embeddings, selecting an appropriate prefix dimension (ranging from 128 to 1536) based on the specific surface and use case. A short-term cache is also employed to handle popular queries and reduce redundant computation. Our online service is thoroughly instrumented with dashboards to monitor throughput, latency, and error rates, allowing for safe canary deployments and quick rollbacks~\cite{covington2016youtube,huang2020facebook}.

\subsection{Offline Inference and Indexing}
Given the scale of our document corpus, offline inference is critical. To manage costs, we embed documents at the feature level (e.g., store or item) and then join these embeddings back to the full catalog of billions of candidates before indexing. The entire process is automated and scheduled on a bi-weekly cadence, which includes a full model retraining and index rebuild. To ensure freshness, we run delta pipelines on a regular cadence to embed new or updated documents. The resulting embeddings are stored in stable feature store tables, keyed by entity IDs, making them readily available for both retrieval and ranking models. These embeddings are used to build our Uber Internal \textbf{Lucene Plus} Indices, which are supported by \textbf{HNSW}~\cite{malkov2018hnsw}  graphs and store both non-quantized (float32) and quantized (int8) vector representations to optimize for various latency and storage budgets~\cite{jegou2011pq,guo2020anisotropic,guo2020scann}. We will provide a dedicated engineering blog for the details.

\section{Methods: Data, Model, Losses, and Training}
This section details the core components of our semantic retrieval system, from data collection and feature engineering to model architecture and training methodology.

\subsection{Data and Features}
Our training data is derived from large-scale implicit user feedback within Uber Eats search logs, spanning from December 2023 to March 2025. This results in approximately several hundreds of millions of positive (query, document) pairs, which form the basis for our domain adaptation phase. Such implicit feedback pipelines have been widely adopted in retrieval systems for e-commerce and recommendations~\cite{covington2016youtube,huang2020facebook,lu2020twinbert}.  We leverage diverse signals from different verticals to capture a comprehensive view of user intent:
\begin{itemize}[leftmargin=*]
\item \textbf{Global Search (GS):} Positive pairs are defined by (query, store) clicks.
\item \textbf{In-Store Search---Online Food Delivery (ISS-OFD):} Positive pairs are defined by (query, dish) clicks.
\item \textbf{In-Store Search---Grocery \& Retail (ISS-GR):} Positive pairs are defined by (query, grocery item) ``add-to-cart'' (ATC) events. The ATC signal in GR provides a strong proxy for relevance due to the high-intent nature of the action.
\end{itemize}

To create a balanced training set and prevent the model from overfitting to the most frequent data, we perform a controlled downsampling of the source logs. The final dataset is rebalanced across English/non-English languages, and across verticals: stores, dishes, and grocery items, consistent with data balancing strategies for large-scale contrastive learning~\cite{chen2020simclr,gao2021simcse}. For features, both queries and documents are represented as structured JSON objects, which aligns with the pretraining of the Qwen2 backbone~\cite{qwen2024report,alibaba2024gteqwen}.

\begin{itemize}[leftmargin=*]
\item \textbf{Queries:} The input is a JSON blob containing search query, search location, search language and other realtime contextual text information. This format allows the model to leverage country and language context to resolve ambiguities (e.g., ``pan'' in Spanish vs. English), and supports cross-lingual generalization~\cite{li2023gte,google2025gemini}.  We truncate the max number of tokens for queries to be 128.
\item \textbf{Documents:} Each vertical has its own JSON schema, which includes core fields such as names/titles, categories, and descriptions. For stores, we also include fields like cuisine types, tags, and popularity summaries, similar to structured encoders in multi-vertical retrieval~\cite{huang2020facebook,lu2020twinbert}. We truncate the max number of tokens for all document types to be 1024.
\end{itemize}

The training data pipeline is designed to be flexible, generating a list-centric schema where each row represents a query and contains lists of positive and negative document IDs, corresponding feature blobs, and per-example weights. This universal format allows us to seamlessly switch between InfoNCE and triplet loss objectives without altering the data pipeline.

\subsection{Model Architecture}
\begin{figure}[t]
\centering
\resizebox{0.9\linewidth}{!}{
\begin{tikzpicture}[
  every node/.style={align=center},
  >={Stealth[length=2mm]},
  box/.style={draw, rounded corners, minimum height=8mm, inner sep=3pt},
  q/.style={box, fill=blue!5,  minimum width=26mm},
  d/.style={box, fill=green!5, minimum width=26mm},
  tq/.style={box, fill=blue!10,  minimum width=36mm},
  td/.style={box, fill=green!10, minimum width=36mm},
  eq/.style={box, fill=blue!5,   minimum width=33mm},
  ed/.style={box, fill=green!5,  minimum width=33mm},
  head/.style={box, thick, fill=gray!10, minimum width=48mm, minimum height=12mm}
]

\def\rowgap{14mm}
\def\colgap{12mm}

\node[q]                      (qinput) {Query\\JSON};
\node[tq, right=\colgap of qinput] (qtower) {Query Tower\\(gte--Qwen2)};
\node[eq, right=\colgap of qtower] (qemb)   {Query Embedding};
\node[head, right=\colgap of qemb] (sim)    {Cosine / FFN\\Scoring Head}; 

\node[d,  below=\rowgap of qinput] (dinput) {Document\\JSON};
\node[td, right=\colgap of dinput] (dtower) {Document Tower\\(gte--Qwen2)};
\node[ed, right=\colgap of dtower] (demb)   {Document Embedding};
\node[box, fill=gray!08, text width=52mm, right=\colgap of demb] (mrl)
  {\footnotesize \textbf{MRL loss (multi-dimension)}\\[-1pt]
   \footnotesize InfoNCE $\rightarrow$ Triplet NCE}; 

\draw[->, thick] (qinput) -- (qtower);
\draw[->, thick] (qtower) -- (qemb);
\draw[->, thick] (dinput) -- (dtower);
\draw[->, thick] (dtower) -- (demb);

\draw[->, thick] (qemb.east) -- (sim.west);                 
\draw[->, thick] (demb.east) -- ($(sim.west)!0.55!(sim.south west)$); 
\draw[->, thick] (sim.south) -- ($(sim.south)!0.5!(mrl.north)$) -- (mrl.north); 

\end{tikzpicture}
}
\caption{Two-tower gte-Qwen2 architecture: query and document towers produce embeddings scored by cosine or an FFN head; training uses InfoNCE, Triplet NCE, and MRL.}
\label{fig:two_tower_architecture}
\end{figure}
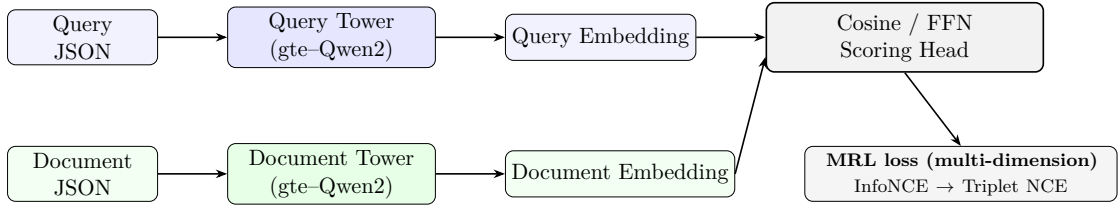

As shown in Figure~\ref{fig:two_tower_architecture}, our model is a two-tower encoder~\cite{huang2013dssm,shen2014cdssm,huang2020facebook} with an unshared weight configuration between the query and document towers, which provides greater capacity and flexibility to learn distinct representations for each. The backbone is the \textbf{gte-Qwen2-1.5B} model, chosen for its strong performance and rich world knowledge.

Inputs are presented to the model as \textbf{JSON-formatted strings}, a strategy that we found to significantly outperform simple text concatenation in our ablations. Each input is encoded by the \textbf{gte-Qwen2-1.5B} backbone~\cite{qwen2024report,alibaba2024gteqwen}, and the final representation is obtained through \textbf{End-of-Sentence (EOS)} token pooling~\cite{karpukhin2020dpr,khattab2020colbert}, which leverages the bidirectional attention mechanism of the transformer to create a single, contextually rich embedding vector. Ablation studies confirmed that EOS pooling consistently outperforms last-token and average pooling approaches, producing more stable cross-lingual and multi-vertical representations. The default similarity metric for retrieval is \textbf{cosine similarity}, which is efficient and consistent between training and inference. In addition, we explore a learned \textbf{feed-forward neural (FFN) distance function} as a non-linear alternative to cosine similarity, designed to recover fine-grained semantic relationships lost in purely linear similarity measures~\cite{zhai2023neuralretrieval,linkedin2024linr}. A dedicated ablation study compares these two scoring methods, quantifying their trade offs in retrieval accuracy, latency, and computational overhead.

\subsection{Objectives and Optimization}
The primary training objective is a multi-stage process designed to produce a robust, multi-fidelity embedding space. Our approach combines two powerful loss functions, InfoNCE~\cite{vandenOord2018cpc,chen2020simclr,gao2021simcse} and a specialized Triplet NCE Loss~\cite{schroff2015facenet}, with \textbf{MRL}~\cite{kusupati2022mrl} to ensure a single model can meet diverse retrieval latency requirements.

For the initial training stage (InfoNCE Mode), we specifically utilize InfoNCE loss with in-batch negatives with a temperature parameter $\tau$ of 0.07 following proven large-batch contrastive training practices~\cite{chen2020simclr,gao2021simcse}, and micro batch size of $512$. This loss is a powerful objective for pushing positive pairs (e.g., a query and a clicked item) closer together while simultaneously pushing them away from all other negative pairs within the same batch.
For a batch of queries $\{q_i\}_{i=1}^N$ and paired positives $\{d_i^+\}_{i=1}^N$, define similarity $s_{ij} = \mathrm{sim}(q_i,d_j)$ (cosine). With temperature $\tau>0$ and optional weights $w_{ij}$, the InfoNCE loss is
\begin{equation}
\mathcal{L}_{\mathrm{InfoNCE}} = - \frac{1}{N}\sum_{i=1}^N \log \frac{\exp(s_{ii}/\tau)}{\sum_{j=1}^N \exp(s_{ij}/\tau)}.
\end{equation}

For a second fine-tuning stage (NCE Mode), we introduce the Triplet NCE Loss objective using hard positive and negative examples for each given query. These challenging examples are programmatically mined from our logs and then verified and labeled by a powerful LLM-based oracle~\cite{google2025gemini,nvidia2024nvembed}. Two additional safety measures are added to ensure the quality of those datasets. First, the LLM-based oracle's performance and accuracy was validated by our engineering team through human review. Second, the examples with LLM labels are further filtered with ctr/cvr rates to remove outliers. This fine-tuning stage helps to sharpen the model's decision boundary and improve retrieval accuracy on the most challenging cases.

For each query $q$, collect a set of positives $\mathcal{P}(q)$ and negatives $\mathcal{N}(q)$ (hard negatives from logs and LLM--verified labels). Let $z_{qd} = \mathrm{sim}(q,d)/\tau$. The Triplet NCE objective combines positive and negative logistic terms:
\begin{align}
\mathcal{L}_{\mathrm{Triplet\mbox{-}NCE}} = \mathbb{E}_{(q,d^+)\sim \mathcal{P}}\left[\log\!\left(1+\exp(-z_{q d^+})\right)\right] + 
\mathbb{E}_{(q,d^-)\sim \mathcal{N}}\left[\log\!\left(1+\exp(+z_{q d^-})\right)\right].
\end{align}

To meet diverse latency requirements across different search surfaces, we integrate Matryoshka Representation Learning (MRL)~\cite{kusupati2022mrl}. This technique allows us to expose nested prefixes of the final 1536-dimensional embedding at various sizes—specifically, at \{128, 256, 512, 768, 1024, 1280, 1536\} dimensions. This provides a single model that can serve a variety of performance-latency trade-offs at inference time. To achieve this, our custom MatryoshkaLoss module wraps the base loss (either InfoNCE or Triplet Loss) and applies it across all nested dimensions. The total loss for a given training step is the sum of the base loss calculated at each of the MRL dimensions:
Let $\mathcal{M}$ be the set of MRL cuts and $E^{(m)}(\cdot)$ the truncation to the first $m$ dimensions. For base loss $\mathcal{L}_\star$ (either InfoNCE or Triplet NCE), the MRL objective is
\begin{equation}
\mathcal{L}_{\mathrm{MRL}} = \sum_{m \in \mathcal{M}} c^{(m)} \, \mathcal{L}_\star\!\left(E^{(m)}(q), E^{(m)}(d)\right),
\end{equation}
with $c^{(m)}\!=\!1$ by default. This aggregation of losses across all prefixes ensures that each nested dimension is independently a semantically meaningful representation, forcing the most critical information to be encoded in the earliest dimensions.

We use the Adam optimizer~\cite{kingma2015adam} with a fixed learning rate of $5\times10^{-5}\,$ without warmup. To preserve the valuable knowledge gained during the gte-Qwen2 pretraining, we only fine-tune the last three layers of the model. This strategy significantly accelerates convergence while preventing "catastrophic forgetting" of the model's general linguistic capabilities~\cite{howard2018ulmfit}. We also explored alternative loss functions, such as the \textbf{SigLIP}-style pairwise sigmoid loss~\cite{zhai2023siglip} (as a contrast to CLIP’s softmax-based objective~\cite{radford2021clip}) to investigate its effect on retrieval performance.

\subsection{Training Setup at Scale}

Our training pipeline runs on Uber’s internal \textbf{Michelangelo Canvas} platform~\cite{uber2025michelangelo}. The model is implemented in PyTorch~\cite{paszke2019pytorch} with Hugging Face Transformers~\cite{wolf2020transformers}; distributed training uses torch.distributed DDP (NCCL) with Ray~\cite{moritz2018ray} for orchestration and DeepSpeed ZeRO-3~\cite{rajbhandari2020zero} to shard optimizer states, gradients, and parameters across multiple GPUs. We employ mixed precision~\cite{micikevicius2018mixedprecision}(bf16 on H100 with fp16 fallback), gradient accumulation to reach large effective batch sizes, activation checkpointing, and periodic checkpointing at fixed step intervals for fault tolerance and reusability. A typical end-to-end run over several hundred of millions query–document pairs completed in 24–48 hours on 8×H100. All experiments are tracked in Comet (loss curves, hit-rate metrics, hyper-parameters), organized via a round/run convention for reproducibility, and all artifacts (checkpoints, configs, evaluation reports) are systematically archived.

\section{Evaluation}

We conducted comprehensive offline and online experiments to validate the performance of our semantic embedding system. During offline evaluations, we evaluated the model’s recall performance on a golden dataset through offline kNN search, following established dense retrieval evaluation protocols~~\cite{karpukhin2020dpr,khattab2020colbert}; during online evaluations, we integrated semantic search into our search stack and compared conversion rates using randomized controlled experiments~\cite{kohavi2009experiments,kohavi2020abtesting}.

\subsection{Models and Variants Compared}
We benchmark our two-tower system along three axes. Baselines include the vanilla \textbf{gte-Qwen2} models prior to any domain tuning and the legacy production \textbf{Mini-LM }\cite{wang2020minilm}, which together establish the gap between modern LLM backbones and a compact, task-specific model. Training regimes compare a single-stage InfoNCE objective~\cite{vandenOord2018cpc,chen2020simclr} with a two-stage recipe that adds Triplet-NCE~\cite{schroff2015facenet} to sharpen hard boundaries; this isolates the incremental value of hard-example supervision beyond large-scale contrastive learning. Serving choices evaluate operational trade-offs: we deploy the fine-tuned model over a quantized ANN index to reduce memory and latency while monitoring quality, and we add a compact non-linear scoring head for micro-re-ranking on the retrieved set to recover fine-grained query–document interactions beyond cosine similarity. Across these conditions, we report retrieval metrics (R@20/200), plus online business KPIs, to attribute gains to backbone capacity, training strategy, and serving configuration.

\subsection{Offline k-NN Retrieval Protocol}

To validate our model's performance under realistic production constraints, the offline evaluation simulates the retrieval process using the following k-NN protocol:
\begin{itemize}[leftmargin=*]
\item \textbf{Precompute Embeddings: } We ensure all document embeddings are pre-computed offline and correspond to the \textit{tte\_id} of the query model being evaluated. Queries are embedded either on the fly or from a pre-computed palette.
\item \textbf{Construct Candidate Sets:} For each (query, geographic location) row from our evaluation dataset, we dynamically generate a comprehensive candidate set. This set includes all stores, dishes, or items that are deliverable within the query's specified hex or city .
\item \textbf{k-NN Evaluation: } The core of the evaluation is a custom k-NN evaluator. For each query, it performs a k-NN search on the pre-constructed candidate set to generate a ranked list of documents with their associated similarity scores with schema query $\rightarrow \, $ [\{ item\_id:  item456, score: 0.9 \},  ... , \{ item\_id:  item123, score: 0.15 \}].
\item \textbf{Metric Aggregation: } We aggregate the results by market and vertical, reporting the mean recall@k at various k values (e.g., k=20, 200, 500, 2k). Recall is defined as the fraction of true positive items found within the top-k retrieved results. For illustrative point of view, we only report recall@20 and recall@200 for offline model evaluation selection, and we do see the consistent trend between recall@200 and recall@500 and recall@2k.   
\end{itemize}

\subsection{Online A/B Testing Protocol}
During online evaluation, the semantic search results are utilized as a dedicated retrieval channel and blended with other results (e.g., lexical/keyword~\cite{robertson2009bm25}) within the search stack. These retrieved results are subsequently fed into the final ranking stage.
The A/B testing protocol measures the end-to-end impact of the new model and its operational variants under live, production traffic, focusing on the system's ability to drive business outcomes.
\begin{itemize}[leftmargin=*]
\item \textbf{Treatment Evaluation: } We specifically test operational trade-offs by comparing variants like the quantized (INT8) versus non-quantized (FP32) index~\cite{jegou2011pq,guo2020anisotropic}.
\item \textbf{Primary Metrics: } The Overall Evaluation Criterion is defined by customer \textbf{Conversion Rate (CVR)} and other metrics like \textbf{Search CoPSU (Orders/Clicks Per Search Unit)}.
\end{itemize}

\subsection{Offline Results: Model Superiority and Training Strategy}

Table~\ref{tab:two-stage} isolates the performance contributions of our architecture and two-stage training strategy against key baselines. The results are presented as the mean Recall across all EN markets (CAN, GBR, US) to reduce the variance of metrics, but the same conclusion holds for all markets.
The final Two-Stage finetuned \textbf{gte-Qwen2} model (InfoNCE + Triplet NCE) delivers a substantial lift across all verticals. We observe a headline improvement of +88\% on high-precision metrics (Store R@20) and 31\% on high-recall metrics (Store R@200) compared to the Legacy Mini-LM baseline, validating the massive retrieval headroom identified in the project's background.
Crucially, the transition from the Single-Stage (InfoNCE Only) to the Two-Stage approach provides an incremental gain of 1−4\% in recall across all verticals. This small, consistent lift validates the necessity and utility of the complex LLM-labeling pipeline and the Triplet NCE objective, confirming that the second stage is effective in sharpening the model's decision boundary for the most challenging, commercially relevant query-document pairs~\cite{google2025gemini,nvidia2024nvembed}.

\begin{table}[t]
\centering
\footnotesize
\setlength{\tabcolsep}{3pt}
\renewcommand{\arraystretch}{1.15}
\caption{finetune QWen2 vs Mini-LLM and vanilla QWen2 for EN countries on different verticals}
\label{tab:model-variants}
\resizebox{\linewidth}{!}{%
\begin{tabular}{L{0.28\linewidth} L{0.18\linewidth} *{6}{C{0.08\linewidth}}}
\toprule
Model Variant & Training Strategy &
\multicolumn{2}{c}{Store} & \multicolumn{2}{c}{Dish} & \multicolumn{2}{c}{GR} \\
& & R@200 & R@20 & R@200 & R@20 & R@200 & R@20 \\
\midrule
finetuned Mini-LM (Baseline) & Single-Stage             & 0.521 & 0.162 & 0.372 & 0.129 & 0.891 & 0.561 \\
Qwen-1.5B (Vanilla)          & No Fine-Tuning           & 0.452 & 0.208 & 0.353 & 0.141 & 0.572 & 0.352 \\
Qwen-1.5B FT (Single-Stage)  & InfoNCE Only             & 0.658 & 0.283 & 0.423 & 0.171 & 0.899 & 0.558 \\
Qwen-1.5B FT (Two-Stage)     & InfoNCE + Triplet NCE    & 0.678 & 0.304 & 0.462 & 0.182 & 0.942 & 0.579 \\
\midrule
\multicolumn{2}{l}{Final Model Lift (vs.\ Legacy)}  & 31\% & 88\% & 24\% & 38\% & 6\%  & 4\%  \\
\multicolumn{2}{l}{Final Model Lift (vs.\ Vanilla)} & 51\% & 42\% & 31\% & 28\% & 65\% & 66\% \\
\bottomrule
\end{tabular}
}
\label{tab:two-stage}
\end{table}

\subsection{Offline Results: Operational Efficiency (MRL, Quantization)}
Table~\ref{tab:mrl_quant} quantifies the impact of our cost and latency optimization strategies. These ablations benchmark the final Two-Stage finetuned gte-Qwen2 model against the uncompressed, full-dimensional standard to validate mechanisms for reducing total cost of retrieval.
The efficacy of \textbf{Matryoshka Representation Learning (MRL)}~\cite{kusupati2022mrl}  is proven by observing that truncating the embedding from 1536D to the 256D cut results in a negligible drop of only 0.002 percentage points in average R@200 (0.678 to 0.676). This near-parity performance is exchanged for a massive reduction: the storage footprint is reduced to approximately 16\% of the full-size vector, and the index cost is cut to 23\% of the original. This validates MRL as a core cost-saving mechanism.
Applying INT8 quantization to the MRL-256 cut in \textbf{Lucene Plus} index further reduces the index size to a mere 4\% and the index cost to 15\% of the original 1536D FP32 vector~\cite{johnson2019faiss,malkov2018hnsw}. . Although this introduces a mild recall penalty (R@200 drops from 0.676 to 0.668) compared to full precision KNN algorithm in offline evaluation, the resulting cost reduction is profound, validating the Final Production Cut as the most operationally efficient configuration for massive-scale retrieval.  

\begin{table}[t]
\centering
\scriptsize
\setlength{\tabcolsep}{4pt}
\caption{Efficiency trade-offs of MRL truncation and INT8 quantization on EN/ES Markets. \\  Index size and cost are shown relative to the 1536-D FP32 baseline. \\}
\label{tab:mrl_quant}
\resizebox{\columnwidth}{!}{%
\begin{tabular}{lccccc}  
\toprule
\textbf{Config} & \textbf{Dim} & \textbf{Type} &
\textbf{Index Size (×)} & \textbf{Index Cost (×)} & \textbf{Avg R@200} \\
\midrule
Full Precision Baseline & 1536 & FP32 & 1.00 & 1.00 & 0.678 \\
MRL Cut                 &  256 & FP32 & 0.16 & 0.23 & 0.676 \\
Final Production Cut    &  256 & INT8 & 0.04 & 0.15 & 0.668 \\
\bottomrule
\end{tabular}}%
\end{table}

\subsection{Offline Results: Multilingual Breakdown (Store R@200)}
Table~\ref{tab:legacy_qwen} specifically addresses our goal of building a unified multilingual system~\cite{li2023gte,google2025gemini} by showing the raw retrieval performance of the Store vertical across all six markets. The Relative Lift demonstrates that the fine-tuned \textbf{gte-Qwen2} model provides substantial gains in all operating regions, with the largest percentage improvements observed in markets with historically lower baseline performance (JPN and TWN)

\begin{table}[h]
\centering
\small
\caption{Finetuned gte-Qwen2 across markets on GS Data (Store R@200).}
\begin{tabular}{lcccccc}
\toprule
Variant & JPN & CAN & GBR & FRA & MEX & TWN \\
\midrule
Legacy Mini-LM & 0.329 & 0.512 & 0.531 & 0.479 & 0.518 & 0.380 \\
Vanilla gte-Qwen2-1.5B & 0.351 & 0.453 & 0.469 & 0.459 & 0.321 & 0.301 \\
Final gte-Qwen2-FT & 0.559 & 0.669 & 0.681 & 0.633 & 0.724 & 0.624 \\
\bottomrule
\end{tabular}
\label{tab:legacy_qwen}
\end{table}

\subsection{Online Results}
The final validation of our system's efficacy and operational trade-offs was conducted through randomized online A/B experiments~\cite{kohavi2009experiments,kohavi2020abtesting}. , measuring user and business impact under live production constraints. The experiments compared the final Two-Stage \textbf{gte-Qwen2-FT} model and its operational variants against the Legacy \textbf{Mini-LM} Control group across all search surfaces. 
Table~\ref{tab:legacy_qwen_online} confirms that the performance gains observed offline translate directly into significant business value. The \textbf{gte-Qwen2-FT} (Two-Stage) model provided a statistically significant lift across all primary metrics.
Primary Metrics Lift: The new model achieved a lift of +0.44\% on Intentful Search CVR and +0.54\% on Search CoPSU (Orders/Clicks Per Search Unit) compared to the Legacy Control. This validates the core project objective and confirms the superior representation quality learned during the two-stage training.
Search Funnel Health: The model significantly reduced the Zero-Result Rate by −69.10\%, demonstrating dramatically improved coverage and robustness, particularly in multi-vertical and multilingual scenarios.
Latency Trade-off: While the latency of the retrieval service itself remained flat (near Δ0 ms), the end-to-end search latency saw an increase of 18\%. This increase was localized entirely within the subsequent ranking stage, which now processes a larger, higher-quality candidate set retrieved by the more powerful semantic model. This increase is a direct operational cost exchanged for improved recall performance.

\paragraph{The Success of Quantization} The results in Table 4 reveal a critical operational finding regarding vector quantization. Comparing the two production cuts (FP32 vs. INT8)~\cite{jegou2011pq,guo2020scann} at the MRL-256 dimension, the INT8 variant achieved a higher CVR lift (+0.44\%) than the FP32 variant (+0.29\%), while keeping retrieval latency unchanged (Δ0 ms) relative to the legacy system. Offline guardrails were set by tuning HNSW fanout/efSearch to closely match exact k-NN behavior, and we observed a mild absolute recall drop at the retrieval step with quantization in offline evaluation. However, in online applications the retrieval results pass through additional filtering and hydration stages; after these stages, our A/B tests showed no measurable recall difference between FP32 and INT8 variants. Consequently, we observed no negative production impact from INT8 quantization. Taken together—higher CVR, flat retrieval latency, and no downstream recall degradation—INT8 at the 256-D MRL cut is a robust, cost-efficient default for large-scale deployment.

\begin{table}[t]
\centering
\scriptsize
\setlength{\tabcolsep}{4pt}
\caption{Online business metric lift for all surfaces (illustrative) with ANN.}
\label{tab:legacy_qwen_online}
\resizebox{\columnwidth}{!}{%
\begin{tabular}{lccccc}
\toprule
\textbf{Online Variant} &
\textbf{Dim (MRL)} &
\textbf{Vector Type} &
$\boldsymbol{\Delta}$\textbf{ Intentful CVR} &
$\boldsymbol{\Delta}$\textbf{ CoPSU} &
\textbf{p95 Latency ($\Delta$ms)} \\
\midrule
Control (Legacy Mini-LM) & 384 & FP32 & -- & -- & -- \\
MRL Trade-off    & 256 & FP32 & +0.29\% & +0.27\% & 0 \\
Quantization (INT8)       & 256 & INT8 & \textbf{+0.44\%} & \textbf{+0.54\%} & 0 \\
\bottomrule
\end{tabular}}
\end{table}

\section{Ablation Studies}
We ran ablations on model/backbone, input representation, pooling, MRL vs fixed FC projection, loss choice (InfoNCE vs SigLIP), and a non-linear distance head. For model-centric factors we used Hit-Rate@k with random in-batch negatives as a fast proxy for recall which is consistent with common recommender and IR practice~\cite{he2017ncf,manning2008ir}. For strategic factors used full recall and operational metrics. For logits matrix $Z\in\mathbb{R}^{B\times M}$ and labels $y\in\{0,\ldots,M-1\}^B$, the hit rate at cut $k$ is
\begin{equation} \label{eq:hit-k}
\mathrm{H@}k = \frac{1}{B}\sum_{i=1}^B \mathbf{1}\{ y_i \in \mathrm{TopK}_k(Z_{i})\}.
\end{equation}
For MRL, we compute H@$k$ per dimension (e.g., 256D, 512D, 1536D) ~\cite{kusupati2022mrl}.

\subsection*{A. Base Model Capacity: Public vs. Fine-Tuned}
This ablation isolates (i) the benefit of using a modern, large language model (LLM) as a backbone and (ii) the incremental gain from domain adaptation. As shown in Table~\ref{tab:ab_model_variant}, vanilla  \textbf{gte-Qwen2} models (1.5B/7B) already exceed the fine-tuned Mini-LLM baseline on H@10, reflecting stronger world knowledge and multilingual capacity ~\cite{alibaba2024gteqwen,li2023gte,google2025gemini}. Crucially, after domain adaptation,  \textbf{gte-Qwen2-1.5B FT} delivers a large improvement in H@1—1.94× over its own vanilla counterpart (0.351 vs. 0.181), i.e., a +94\% relative lift—confirming that deep domain fine-tuning is essential for our setting ~\cite{karpukhin2020dpr,huang2020facebook}. We also observe that vanilla  \textbf{gte-Qwen2-1.5B} is close to vanilla  \textbf{gte-Qwen2-7B} while being substantially cheaper to train/serve; consequently, we select  \textbf{gte-Qwen2-1.5B} as the production backbone for the first release.

\begin{table}[h]
\centering
\small
\caption{Vanilla vs Finetuned Qwen vs Mini-LLM on GS data.}
\begin{tabular}{lcccc}
\toprule
Models & \#Params & Embed dim & H@1 & H@10 \\
\midrule
Finetuned Mini-LLM & 0.12B & 384 & 0.072 & 0.313 \\
Vanilla gte-Qwen2-1.5B & 1.4B & 1536 & 0.181 & 0.492 \\
Vanilla gte-Qwen2-7B & 7.07B & 3584 & 0.208 & 0.529 \\
Vanilla Qwen3-Embedding-8B  & 7.57B & 4096 & 0.159 & 0.449 \\
Finetuned gte-Qwen2-1.5B & 1.4B & 1536 & 0.351 & 0.832 \\
\bottomrule
\end{tabular}
\label{tab:ab_model_variant}
\end{table}

\subsection*{B. Input Representation: JSON vs Plain}
This experiment addresses the importance of structured input for semantic encoding in a multi-vertical environment. Table~\ref{tab:ab_feature_json} demonstrates that structuring the inputs as JSON fields yields higher H@K scores across the board compared to simple plain text concatenation. On vanilla \textbf{gte-Qwen2-1.5B}, JSON yields absolute gains of 0.023 H@1, 0.031 H@5, and 0.044 H@10 (≈ 14.6\%, 8.8\%, 9.8\% relative), and this suggests that the gte-Qwen2 backbone, which was pre-trained on structured data formats, effectively utilizes the explicit field names (e.g., "store\_name", "cuisine\_type") to disambiguate information, resulting in a cleaner embedding space.

One additional benefit is  uniform multi-vertical support because the json format naturally supports different inputs from different verticals (store/dish/item). We can write them in similar json formats, but with different field names. The model will automatically recognize it is a store/dish/item based on the field name and contents. This enables us to train a single model to embed documents from different verticals via type-aware semantics without bespoke encoders.
\begin{table}[h]
\centering
\small
\caption{Plain text vs JSON input on vanilla gte-Qwen2-1.5B (GS data).}
\begin{tabular}{lccc}
\toprule
Features & H@1 & H@5 & H@10 \\
\midrule
JSON text & 0.181 & 0.383 & 0.492 \\
Plain text & 0.158 & 0.352 & 0.448 \\
\bottomrule
\end{tabular}
\label{tab:ab_feature_json}
\end{table}

\subsection*{C. Pooling Strategy: EOS vs Average vs Last-Token}
We compare aggregation mechanisms that compress token-level representations into a single vector for retrieval. As shown in Table~\ref{tab:ab_pooling}, EOS pooling decisively outperforms both last-token and average pooling—yielding roughly a 6x higher H@1 than last-token and an order-of-magnitude gain over average. This aligns with the backbone’s pretraining: the EOS token functions as a learned summary point that attends over the full prefix, producing a representation that better captures global semantics. In contrast, last-token is anchored to the trailing field or punctuation (especially in JSON inputs) and average pooling dilutes salient fields with boilerplate and separators. We observe the same ranking (EOS » last-token ≥ average) across markets and MRL cuts.
\begin{table}[h]
\centering
\small
\caption{EOS vs last word embedding pooling on GS data.}
\begin{tabular}{lccc}
\toprule
Pooling & H@1 & H@5 & H@10 \\
\midrule
Last word & 0.030 & 0.089 & 0.131 \\
Average & 0.021 & 0.071 & 0.118 \\
EOS & 0.181 & 0.383 & 0.492 \\
\bottomrule
\end{tabular}
\label{tab:ab_pooling}
\end{table}

\subsection*{D. Representation Size \& Method: MRL vs Fixed FC}
This ablation validates Matryoshka Representation Learning (MRL) ~\cite{kusupati2022mrl} as the superior strategy for dimensional trade-offs, focusing on operational flexibility rather than just marginal recall gain. Table~\ref{tab:ab_rml} confirms that MRL cuts maintain recall quality close to the full-dimensional vector and are competitive with a dedicated Fixed FC Head model at the same truncated dimension. While the end-to-end recall performance for the 192D and 256D cuts is statistically similar to the corresponding Fixed FC projection models, MRL offers significant architectural and operational advantages:
\begin{itemize}[leftmargin=*]
\item \textbf{Dynamic Budgeting:} MRL enables a single deployed model to instantly deliver embeddings at any supported dimension (e.g., {128,256,…,1536}). In contrast, the Fixed FC approach requires deploying and maintaining a new model variant for every desired dimension, leading to massive operational overhead.
\item \textbf{Full Vector Retention:}  MRL preserves access to the original, high-quality 1536D embedding for use in downstream ranking models. The Fixed FC Projection permanently restricts the vector size, losing this richer, higher-dimensional context.
\item \textbf{Simplified Maintenance:} By consolidating multiple vector sizes into one trained model, MRL significantly reduces the deployment, monitoring, and index management complexity.
\end{itemize}

\begin{table}[h]
\centering
\small
\caption{Recall for different MRL cuts on GS Data.}
\begin{tabular}{lccc}
\toprule
Dimension Cut & EN R@200 & TW R@200 & ES R@200 \\
\midrule
MRL-192 & 0.671 & 0.613 & 0.719 \\
MRL-256 & 0.676 & 0.619 & 0.722 \\
MRL-1536 & 0.678 & 0.624 & 0.724 \\
FC-192 & 0.676 & 0.614 & 0.717 \\
\bottomrule
\end{tabular}
\label{tab:ab_rml}
\end{table}

\subsection*{E. Loss: InfoNCE vs SigLIP}
We evaluated SigLIP as an alternative to the standard InfoNCE loss to explore non-softmax contrastive training. As shown in Table~\ref{tab:ab_siglip}
, SigLIP~\cite{zhai2023siglip} exhibits a slightly better performance ceiling, particularly on the higher-recall metric (R@200), and shows better stability across varying batch sizes compared to InfoNCE~\cite{vandenOord2018cpc,chen2020simclr}.
A critical finding of this ablation relates to batch sensitivity: the InfoNCE loss experienced a noticeable degradation in Recall@20 as batch size decreased (from 0.280 at B=384 to 0.263 at B=64). In contrast, the SigLIP loss maintained higher performance across the same range, demonstrating its robustness against smaller batch sizes. Specifically, SigLIP at batch size 128 (R@200=0.666) outperforms InfoNCE at the largest batch size of 384 (R@200=0.658). This suggests that SigLIP is less constrained by the effective batch size than InfoNCE, making it a viable or slightly superior alternative for scaling up the initial domain adaptation phase, especially when GPU memory is a bottleneck. 
However, for the initial production deployment, we selected InfoNCE Loss due to its wider adoption, simpler configuration, and established convergence stability with our existing infrastructure and large batch sizes. We plan to explore the integration of SigLIP further, particularly in the future by replacing InfoNCE in the first stage and combining it with the second-stage Triplet NCE Loss ~\cite{schroff2015facenet} to maximize the performance ceiling.

\begin{table}[h]
\centering
\small
\caption{Recall for InfoNCE vs SigLIP on GS Data.}
\begin{tabular}{lccccccc}
\toprule
\textbf{Loss} & \textbf{Batch size} & \textbf{Init.\ temp.} & \textbf{Init.\ bias} & \textbf{R@20} & \textbf{R@200} & \textbf{R@500} \\
\midrule
InfoNCE & 512 & 0.07 & NA & 0.282 & 0.659 & 0.806 \\
InfoNCE & 384 & 0.07 & NA & 0.280 & 0.658 & 0.807 \\
InfoNCE & 192 & 0.07 & NA & 0.274 & 0.651 & 0.803 \\
InfoNCE & 128 & 0.07 & NA & 0.274 & 0.652 & 0.803 \\
InfoNCE &  64 & 0.07 & NA & 0.263 & 0.639 & 0.795 \\
\midrule
SigLip  & 512 & 0.03 & $-6$ & 0.294 & 0.648 & 0.795 \\
SigLip  & 384 & 0.03 & $-6$ & \textbf{0.303} & 0.664 & 0.808 \\
SigLip  & 192 & 0.06 & $-6$ & 0.293 & 0.662 & 0.810 \\
SigLip  & 128 & 0.10 & $-6$ & 0.291 & \textbf{0.666} & \textbf{0.812} \\
SigLip  &  64 & 0.10 & $-6$ & 0.286 & 0.660 & 0.806 \\
\bottomrule
\end{tabular}
\label{tab:ab_siglip}
\end{table}

\subsection*{F. Non-linear Distance Function (Micro Re-ranking)}


This ablation quantifies the gain achieved by introducing a compact learned interaction layer 
(\textit{Neural Distance Function}) for micro-re-ranking near the retrieval stage, 
which serves to recover fine-grained relevance lost by the fixed cosine similarity function~\cite{zhai2023neuralretrieval,linkedin2024linr}. 
The retrieval process itself involves two sequential steps: 
first, retrieving $10\times$ top-$k$ candidates using the fast dot-product score derived directly 
from the two-tower embeddings, and second, applying the non-linear distance function 
to re-rank and select the final top-$k$ results.

The neural scoring function’s architecture $g(\mathbf{q}, \mathbf{d})$ is implemented 
as a lightweight feed-forward network (FFN) that takes the concatenation of the query 
and document embeddings as input:
\begin{equation}\label{eq:neural-head}
\mathbf{z} = [\mathbf{q}; \mathbf{d}], \qquad 
s = g(\mathbf{q}, \mathbf{d}) = \text{FFN}(\mathbf{z}),
\end{equation}
where $\text{FFN}(\cdot)$ denotes a three-layer network composed of linear projections, 
GELU activations, Layer Normalization, and dropout regularization:
\begin{align}
\mathbf{h}_1 &= \text{GELU}(W_1 \mathbf{z} + b_1), \\
\mathbf{h}_2 &= \text{LayerNorm}(W_2 \mathbf{h}_1 + b_2), \\
s &= W_3 \mathbf{h}_2 + b_3.
\end{align}
Optionally, a $\tanh$ activation is applied to bound the final score. This design directly models the joint interaction of $\mathbf{q}$ and $\mathbf{d}$  through concatenation rather than explicit element-wise products,  yielding a symmetric, GPU-efficient scoring head.

For consistency with the two-tower encoders, the neural distance model  is trained using the same two-stage procedure:  an initial \textbf{InfoNCE} phase with in-batch negatives for large-scale domain adaptation~\cite{vandenOord2018cpc,chen2020simclr},  followed by a \textbf{Triplet NCE}~\cite{schroff2015facenet}  fine-tuning stage that leverages mined hard positives  and negatives from the same dataset used for the original model. 
The Qwen2 encoders remain frozen during this process to preserve the stability of  the base embedding space. As shown in Table~\ref{tab:ffn_recall200} and Table~\ref{tab:ffn_recall500},  the FFN-based non-linear scoring head provides a small but consistent relative lift  of approximately $+1$–$2$\% in Recall@200 and Recall@500 across all markets  compared to cosine similarity.  These results demonstrate that even a compact neural scoring layer can recover fine-grained semantic relevance that the fixed cosine function fails to capture.  They also motivate future exploration of more expressive, coupled architectures,  such as jointly fine-tuning the scoring head with the encoders or adopting  mixture-of-logits~\cite{zhai2023neuralretrieval} similarity functions to further enhance ranking fidelity.

\begin{table}[htbp]
\centering
\scriptsize
\setlength{\tabcolsep}{4pt}
\begin{tabular}{lcccccccc}
\toprule
\textbf{Model Variant} & \textbf{Metric} & \textbf{JPN} & \textbf{CAN} & \textbf{GBR} &
\textbf{FRA} & \textbf{MEX} & \textbf{TWN} & \textbf{USA} \\
\midrule
2-Stage Finetune Qwen2 & Recall@200 & 0.559 & 0.669 & 0.681 & 0.633 & 0.724 & 0.624 & 0.678 \\
\shortstack[l]{Frozen Final Qwen2\\+ Non-linear distance} & Recall@200 & 0.572 & 0.668 & 0.693 & 0.634 & 0.751 & 0.630 & 0.690 \\
\midrule
\textit{Lift (\%)} &  & 2.33 & –0.15 & 1.76 & 0.16 & 3.73 & 0.96 & 1.77 \\
\bottomrule
\end{tabular}
\caption{Recall@200 of dot-product vs.\ FFN scoring on embeddings (\textbf{GS} data).}
\label{tab:ffn_recall200}
\end{table}

\begin{table}[htbp]
\centering
\scriptsize
\setlength{\tabcolsep}{4pt}
\begin{tabular}{lcccccccc}
\toprule
\textbf{Model Variant} & \textbf{Metric} & \textbf{JPN} & \textbf{CAN} & \textbf{GBR} &
\textbf{FRA} & \textbf{MEX} & \textbf{TWN} & \textbf{USA} \\
\midrule
2-Stage Finetune Qwen2 & Recall@500 & 0.691 & 0.818 & 0.828 & 0.756 & 0.848 & 0.744 & 0.807 \\
\shortstack[l]{Frozen Final Qwen2\\+ Non-linear distance} & Recall@500 & 0.697 & 0.831 & 0.841 & 0.759 & 0.864 & 0.751 & 0.822 \\
\midrule
\textit{Lift (\%)} &  & 0.95 & 1.57 & 1.66 & 0.39 & 1.95 & 1.00 & 1.86 \\
\bottomrule
\end{tabular}
\caption{Recall@500 of dot-product vs.\ FFN scoring on embeddings (\textbf{GS} data).}
\label{tab:ffn_recall500}
\end{table}

\section{Conclusion and Future Work}
We presented a production-oriented embedding system for Uber Eats that provides a unified, multilingual, and multi-vertical solution for search retrieval. By leveraging a two-tower Qwen2 backbone~\cite{huang2020facebook,alibaba2024gteqwen,qwen2024report} and fine-tuning it with a robust InfoNCE objective, we achieved substantial recall gains over our baseline across six diverse markets while meeting strict latency and operational constraints. Our system incorporates Matryoshka Representation Learning (MRL)~\cite{kusupati2022mrl} to serve multiple embedding dimensions from a single model, offering a flexible trade-off between quality and latency.

The operational design is a key contribution of this work. We seamlessly integrated our model with Lucene Plus through per-vertical indices, market sharding, and effective pre-filters~\cite{robertson2009bm25,huang2020facebook}. The use of a blue/green column strategy with a configuration ensures safe, atomic cutovers that maintain consistency between online query models and offline indices~\cite{kohavi2009experiments,kohavi2020abtesting}. Furthermore, we optimize for storage and serving costs by indexing both non-quantized and quantized embeddings~\cite{malkov2018hnsw,johnson2019faiss,jegou2011pq,guo2020scann,guo2020anisotropic}. A fully automated, bi-weekly auto-retrain $\rightarrow$ auto-deploy $\rightarrow$ auto-index pipeline, supplemented by daily deltas and rigorous gateway validation, guarantees freshness, stability, and safety~\cite{covington2016youtube,huang2020facebook}.
Empirically, our findings show that domain adaptation on large-scale, implicit user data is the primary driver of performance improvements~\cite{covington2016youtube,karpukhin2020dpr}. The use of JSON inputs and EOS pooling consistently proved to be the most robust choices for our multilingual and multi-vertical domain~\cite{lu2020twinbert,karpukhin2020dpr,khattab2020colbert}. We also demonstrated that MRL provides a superior alternative to managing multiple fixed-size models, and that a lightweight neural distance head can provide a significant boost to low-k recall with minimal latency overhead~\cite{zhai2023neuralretrieval,linkedin2024linr}.

Despite its effectiveness, our system has several limitations. While our online A/B testing captures real-world user behavior and business impact, a controlled experiment cannot fully capture the long-term impact on user retention~\cite{kohavi2009experiments}. Some cross-vertical false negatives inevitably remain due to the nature of in-batch negative sampling on our large, heterogeneous dataset. While our neural distance head shows promise, its use must be carefully budgeted to avoid violating latency SLOs. Furthermore, while our blue/green column strategy is highly effective, it is not as isolated or flexible as the long-term goal of using fully independent indices for each model version. Our roadmap for future development focuses on four key areas:
\begin{itemize}[leftmargin=*]
\item \textbf{Capacity \& Modeling:} We will scale our training to larger backbones (7B/14B) and explore more sophisticated modeling techniques, including late-interaction layers for re-ranking~\cite{khattab2020colbert}. We also plan to refine our neural distance head through better feature engineering and explore advanced mixture loss functions guided by hard-example mining~\cite{schroff2015facenet,settles2009active}. A major area of future work is the incorporation of multi-modal signals, such as images, into our document tower ~\cite{radford2021clip,zhai2023siglip}to capture richer item representations.
\item \textbf{Data \& Labeling:} We will enhance our LLM-assisted relevance curation pipeline to include active learning loops, strengthen our hard negative mining across all languages and verticals, and introduce proactive drift monitors to detect performance degradations in specific markets.
\item \textbf{Production Search:} Our north star is to move towards a fully isolated-index rollout per model run, which will require building on-demand swing capacity. We will also continue our work on quantization and compaction techniques to reduce index footprint and latency, and automate the tuning of ANN parameters and MRL dimensions for different surfaces~\cite{johnson2019faiss,malkov2018hnsw,jegou2011pq}.
\item \textbf{Unified Objectives:} We will advance our training by moving to a multi-objective learning framework that jointly optimizes for both retrieval relevance and downstream pCVR, with explicit constraints for fairness and geo-eligibility. Finally, we aim to formalize our entire lifecycle management process to eliminate residual manual steps and further harden our operational guardrails.
\end{itemize}

\section*{Acknowledgments}
We thank collaborators and colleagues (Lei Shi, Frank Zhang, Karthikeyan Ramasamy, Shubham Gupta, Yupeng Fu, Peng Yang, Zhitao Li, Viv Keswani, Chris Harris) for invaluable discussions, engineering support, and feedbacks. 

\clearpage
\bibliographystyle{unsrtnat} 
\bibliography{uber_two_tower_refs}

\end{document}